%Paper: hep-ph/9311304
%From: cho@theory3.caltech.edu (Peter Cho)
%Date: Wed, 17 Nov 93 10:57:38 PST
%Date (revised): Tue, 1 Feb 94 14:32:26 PST

% ---------------------------------------------------------------------
% This file uses the Harvmac macros and should be printed out in
% "big" format.
% ---------------------------------------------------------------------

\input harvmac

% The following command eliminates black boxes that appear when
% equations run over RHS margins:
\overfullrule=0pt

% Small capital subscripts

\def\D{{\scriptscriptstyle D}}

\def\H{{\scriptscriptstyle H}}
\def\I{{\scriptscriptstyle I}}

\def\K{{\scriptscriptstyle K}}
\def\L{{\scriptscriptstyle L}}

\def\P{{\scriptscriptstyle P}}
\def\Q{{\scriptscriptstyle Q}}
\def\R{{\scriptscriptstyle R}}
\def\S{{\scriptscriptstyle S}}
\def\T{{\scriptscriptstyle T}}

% Calligraphy letters

\def\CL{{\cal L}}

% Greek letters

\def\a{\alpha}
\def\b{\beta}
\def\d{\delta}
\def\e{\epsilon}
\def\g{\gamma}

\def\th{\theta}
\def\u{\mu}
\def\v{\nu}

%  Aliases

\def\bar#1{\overline{#1}}
\def\Br{{\rm Br}}
\def\ccdot{\hbox{\kern-.1em$\cdot$\kern-.1em}}

 		% Hyphen in equations

\def\DM{{\Delta M}}

\def\gfive{\gamma^5}
\def\gtap{\raise.3ex\hbox{$>$\kern-.75em\lower1ex\hbox{$\sim$}}}

\def\ltap{\raise.3ex\hbox{$<$\kern-.75em\lower1ex\hbox{$\sim$}}}
\def\LX{{\Lambda_\chi}}

\def\mc{m_c}

\def\MeV{\> {\rm MeV}}
\def\mpi{m_{\pi}}

\def\mQ{m_\Q}

\def\proj{{1 + \slash{v} \over 2}}
\def\projminus{{1 - \slash{v} \over 2}}

\def\slash#1{#1\hskip-0.5em /}

\def\sqrroot{\sqrt{(E_1^2-\mpi^2)(E_2^2-\mpi^2)}}

% Fractions

\def\half{{1 \over 2}}
\def\ninth{{1 \over 9}}
\def\quarter{{1 \over 4}}
\def\sixth{{ 1\over 6}}
\def\third{{1 \over 3}}

\def\twothirds{{2 \over 3}}

% Bold Greek letter macro:

% Style-sensitive Poor-Man's-Bold command, produces bold greek letters.
% Usage $ ... \pmb\gamma ... $
% Adapted from TeXbook p386 (\pmb) and p360 (\mathpallette)
\newdimen\pmboffset
\pmboffset 0.022em
\def\oldpmb#1{\setbox0=\hbox{#1}%
 \copy0\kern-\wd0
 \kern\pmboffset\raise 1.732\pmboffset\copy0\kern-\wd0
 \kern\pmboffset\box0}
\def\pmb#1{\mathchoice{\oldpmb{$\displaystyle#1$}}{\oldpmb{$\textstyle#1$}}
      {\oldpmb{$\scriptstyle#1$}}{\oldpmb{$\scriptscriptstyle#1$}}}

\def\pib{{\pmb{\pi}}}

% Modified title definition to allow long titles to be broken up into three
% lines.

%
% Modified Appendix definition with no unwanted letter after the boldface
% word APPENDIX.
%
\def\appendix#1#2{\global\meqno=1\global\subsecno=0\xdef\secsym{\hbox{#1.}}
\bigbreak\bigskip\noindent{\bf Appendix. #2}\message{(#1. #2)}
\writetoca{Appendix {#1.} {#2}}\par\nobreak\medskip\nobreak}

% ----------------------------------------------------------------------
% References
% ----------------------------------------------------------------------

\nref\WiseI{M. Wise, Phys. Rev. {\bf D45} (1992) R2188.}
\nref\Burdman{G. Burdman and J. Donoghue, Phys. Lett. {\bf B280} (1992) 287.}
\nref\Yan{T.M. Yan, H.Y. Cheng, C.Y. Cheung, G.L. Lin, Y.C. Lin and H.L.
  Yu, Phys. Rev. {\bf D46} (1992) 1148.}
\nref\ChoI{P. Cho, Phys. Lett. {\bf B285} (1992) 145.}
\nref\WiseII{For a general review of Heavy Hadron Chiral Perturbation
 Theory, see M. Wise, Caltech Preprint CALT-68-1860 (1993), Lectures given
 at the CCAST Symposium.}
\nref\CLEO{J.P. Alexander {\it et al.} (CLEO Collaboration), Phys. Lett.
 {\bf B303} (1993) 377.}
\nref\CLEOmesonref{``$D^{**0}$ Production and Decay'', CLEO CONF 93-14,
 contributed to the International Symposium on Lepton and Photon Interactions,
 Ithaca, 1993.}
\nref\CLEObaryonrefs{
 D. Acosta {\it et al.} (CLEO Collaboration), ``Observation of the Excited
 Charmed Baryon $\Lambda_c^*$ in CLEO-II'', CLEO CONF 93-7, contributed to the
 International Symposium on Lepton and Photon Interactions, Ithaca, 1993\semi
 M. Battle {\it et al.} (CLEO Collaboration), ``Observation of a New Excited
 Charmed Baryon'', CLEO CONF 93-32, contributed to the International Symposium
 on Lepton and Photon Interactions, Ithaca, 1993.}
\nref\Fermigrp{P.L. Frabetti {\it et al.} (E687 Collaboration), Fermilab
 preprint Pub-93/249-E (1993).}
\nref\FalkLuke{A. Falk and M. Luke, Phys. Lett. {\bf B292} (1992) 119.}
\nref\Kilian{U. Kilian, J.G. K\"orner and D. Pirjol, Phys. Lett. {\bf B288}
 (1992) 360.}
\nref\GeorgiI{H. Georgi, Heavy Quark Effective Field Theory, {\it in} Proc.
  of the Theoretical Advanced Study Institute 1991, ed. R.K. Ellis, C.T. Hill
  and J.D. Lykken (World Scientific, Singapore, 1992) p. 589.}
\nref\Falk{A. Falk, Nucl. Phys. {\bf B378} (1992) 79.}
\nref\LukeManohar{M. Luke and A. Manohar, Phys. Lett. {\bf B286} (1992)
 348.}
\nref\IsgurWise{N. Isgur and M. Wise, Phys. Rev. Lett. {\bf 66} (1991)
 1130.}
\nref\Lu{M. Lu, M. Wise and N. Isgur, Phys. Rev. {\bf 45} (1992) 1553.}
\nref\QMrefs{E. Eichten, C.T. Hill and C. Quigg, Fermilab preprint
 Pub-93/255-T, hep-ph 9308337\semi
 S. Godfrey and R. Kokoski, Phys. Rev. {\bf D43} (1991) 1679\semi
 J. Rosner, Comments. Nucl. Part. Phys. {\bf 16} (1986) 109, and references
 therein\semi
 S. Godfrey and N. Isgur, Phys. Rev. {\bf D32} (1985) 189.}
\nref\ChoWise{P. Cho and M. Wise, Caltech Preprint CALT-68-1914 (1994).}
\nref\GasserLeutwyler{J. Gasser and H. Leutwyler, Nucl. Phys. {\bf B250}
(1985) 465.}
\nref\Jackson{J. D. Jackson, Introduction to Dispersion Relations, {\it in}
 Dispersion Relations, Scottish Universities' Summer School 1960, ed. G. R.
 Screaton (Oliver and Boyd, 1961) p. 54.}
\nref\CLEODS{Y. Kubota {\it et al.} (CLEO Collaboration), Cornell Preprint
 CLNS 94/1266 (1994).}

% ----------------------------------------------------------------------
% Figure captions
% ----------------------------------------------------------------------

\nfig\Dstwowidth{Total predicted width of $D_{s2}$ plotted as a function of
its mass.  The dashed curves lying above and below the solid curve delineate
the one-standard deviation region about the width's central value.
Additional uncertainties due to $SU(3)$ violation are not pictured.}

\nfig\Gammaplot{Single kaon, single pion and double pion $D'_{s1}$ decay rates
plotted as a function of the heavy meson's mass.  The dashed curve denotes
$\Gamma(D'_{s1} \to D^{*0} K^+) + \Gamma(D'_{s1} \to D^{*+} K^0)$, while the
dotdashed curve corresponds to $\Gamma(D'_{s1} \to D_s^* \pi^0)$.  The
solid curve represents the sum of the widths
$\Gamma(D'_{s1} \to D_s \sum_i \pi^i \pi^i) +
\Gamma(D'_{s1} \to D^*_s \sum_i \pi^i \pi^i)$.  We have assumed the
dimensionless coupling constant values $f_1=1.0$ and $h_1=0.1$ in this
plot.}

% ----------------------------------------------------------------------
% Title page
% ----------------------------------------------------------------------

\def\CITTitle#1#2#3{\nopagenumbers\abstractfont
\hsize=\hstitle\rightline{#1}
\vskip 0.6in\centerline{\titlefont #2} \centerline{\titlefont #3}
\abstractfont\vskip .5in\pageno=0}
\CITTitle{{\baselineskip=12pt plus 1pt minus 1pt
  \vbox{\hbox{CALT-68-1902}\hbox{DOE RESEARCH AND}\hbox{DEVELOPMENT
  REPORT}}}} {Strong Decays of Strange Charmed P-Wave Mesons}{}
\centerline{
  Peter Cho\footnote{$^\dagger$}{Work supported in part by an
  SSC Fellowship and by the U.S. Dept. of Energy under DOE Grant no.
  DE-FG03-92-ER40701.}
  and Sandip P. Trivedi\footnote{$^\ddagger$}{Work supported in part by a
  McCone Fellowship and by the U.S. Dept. of Energy under DOE Grant no.
  DE-FG03-92-ER40701.}}

\centerline{Lauritsen Laboratory}
\centerline{California Institute of Technology}
\centerline{Pasadena, CA  91125}

\vskip .3in
\centerline{\bf Abstract}
\bigskip

	Goldstone boson decays of P-wave $D_s^{**}$ mesons are studied within
the framework of Heavy Hadron Chiral Perturbation Theory.  We first analyze the
simplest single kaon decays of these strange charmed mesons.   We derive a
model independent prediction for the width of $D_{s2}$ and use experimental
information on $D_{s1}$ to constrain the S-wave contribution to $D_1^0$
decay.   Single and double pion decay modes are then discussed and shown to be
significantly restricted by isospin conservation.   We conclude that the
pion channels may offer the best hope for detecting one strange member of an
otherwise invisible P-wave flavor multiplet.

\Date{1/94}

% ----------------------------------------------------------------------
\newsec{Introduction}
% ----------------------------------------------------------------------

	During the past few years, a synthesis of Chiral Perturbation Theory
and the Heavy Quark Effective Theory (HQET) has been developed
\refs{\WiseI{--}\WiseII}.   This hybrid effective theory describes the
low energy strong interactions between light Goldstone bosons and hadrons
containing a heavy quark.  Heavy Hadron Chiral Perturbation Theory (HHCPT)
has been most widely applied to processes involving charm and bottom hadrons
that correspond to ground state mesons and baryons in the nonrelativistic quark
model.  It is however straightforward to incorporate orbital or radial
excitations into the formalism and to study their transitions as well.
Experimental information on such excited mesons and baryons is at present much
less plentiful than for the lowest lying heavy hadrons.  However, data on
excited charm hadrons is currently being collected at CLEO
\refs{\CLEO{--}\CLEObaryonrefs} and Fermilab \Fermigrp, and future
experiments are expected to fill in many of their basic properties.
The phenomenology of these new particles will provide valuable tests of
several basic HQET ideas.

	Of all the possible excited heavy hadrons, P-wave mesons are among
the simplest.  They are characterized in the quark model as heavy quark-light
antiquark bound states carrying one unit of orbital angular momentum.  Such
mesons have been included into the heavy hadron chiral Lagrangian in
refs.~\FalkLuke\ and \Kilian.  The Lagrangian was then used to study single
and double pion decays of ${D^0}^{**}$ and ${D^+}^{**}$ for which the
greatest amount of experimental data is currently available.  In this article,
we would like to extend these previous investigations and focus instead upon
${D_s}^{**}$ states.  As we shall see, isospin considerations significantly
restrict the decays of excited strange charmed mesons and can lead to
qualitatively different results.

	Our paper is organized as follows.  In section~2, we review the
incorporation of P-wave mesons into HHCPT.  We then focus in section~3 upon
$D_s^{**}$ states and discuss their single kaon decay modes.  We derive
a model independent prediction for the $D_{s2}$ width, and we use
experimental limits on $D_{s1}$ to constrain the S-wave component in
$D_1^0$ decays.   We then investigate single and double pion decay
channels in section~4 and discuss prospects for finding a particular
$D_s^{**}$ state that belongs to a class of P-wave mesons which has never
been seen.  Finally, we close in section~5 with a summary of our results.

% ----------------------------------------------------------------------
\newsec{The Heavy Meson Chiral Lagrangian}
% ----------------------------------------------------------------------

	We begin by recalling the fields which enter into the
heavy meson chiral Lagrangian.  The Goldstone bosons resulting from the
chiral symmetry breakdown $SU(3)_\L \times SU(3)_\R \to SU(3)_{\L+\R}$ appear
in the pseudoscalar meson octet
\eqn\pionoctet{\pib = \sum_{a=1}^8 \pi^a T^a = {1 \over \sqrt{2}}
\pmatrix{ \sqrt{\half} \pi^0 + \sqrt{\sixth} \eta & \pi^+ & K^+ \cr
\pi^- & - \sqrt{\half} \pi^0+\sqrt{\sixth}\eta & K^0 \cr
K^- & \bar{K}^0 & - \sqrt{\twothirds}\eta \cr}}
and are conventionally arranged into the exponentiated matrix functions
$\Sigma=e^{2i \pib/f}$ and $\xi=e^{i \pib/f}$.  These matrix functions
transform under the chiral symmetry group
as
\eqn\Sigmafields{\eqalign{\Sigma &\to L \Sigma R^\dagger \cr
\xi &\to L \xi U^\dagger(x) = U(x) \xi R^\dagger \cr}}
where $L$ and $R$ represent global elements of $SU(3)_\L$ and $SU(3)_\R$ while
$U(x)$ acts like a local $SU(3)_{\L+\R}$ transformation.  Chiral invariant
terms describing Goldstone boson self interactions may be constructed from
the fields in \Sigmafields\ and their derivatives.

      Mesons with quark content $Q \bar{q}$ absorb and emit light Goldstone
bosons with no appreciable change in their four-velocities in the infinite
mass limit of their heavy quark constituents $Q$.   They are
consequently represented by velocity dependent fields.  We absorb square
roots of meson masses into these fields' definitions so that one particle
states are normalized as
\eqn\norm{ \vev{p,s|p',s'} = 2 v^0 (2\pi)^3 \d^{(3)}({\vec p}-{\vec p}\,')
\d_{s \, s'}.}
The velocity dependent fields then have mass dimension $3/2$.

	The ground state $J^\P = 0^-$ and $J^\P = 1^-$ mesons which
result from coupling together the heavy quark and light antiquark spins in
an S-wave bound state are annihilated by the pseudoscalar and vector meson
operators $P_i(v)$ and $P^*_{i\u}(v)$.  Their individual components are
given by $(P_1,P_2,P_3)=(D^0, D^+,D_s)$ and
$(P_{1\u}^*,P_{2\u}^*,P_{3\u}^*)=({D^0}^*,{D^+}^*,{D_s}^*)$
when the heavy quark inside the meson is taken to be charm.  In the infinite
quark mass limit, it is useful to combine the degenerate $J^\P=0^-$ and
$J^\P=1^-$ states into the $4\times 4$ matrix field \refs{\WiseI,\GeorgiI}
\eqn\Hfield{H_i(v) = \proj \bigl[ - P_i(v) \gfive + P^*_{i \u}(v) \gamma^\u
\bigr].}
The superfield $H$ carries a heavy quark spinor index and a separate light
antiquark spinor index, and it transforms as an antitriplet under
flavor $SU(3)_{\L+\R}$ and doublet under spin symmetry $SU(2)_v$.

	In P-wave excited mesons, the antiquark spin can pair with
one unit of orbital angular momentum to form states with light angular momentum
$j_\ell = 1/2$ or $j_\ell = 3/2$.  Coupling with the heavy quark spin then
yields two pairs of two degenerate states.  In the first case, the
resulting $J^\P=0^+$ and $J^\P=1^+$ mesons are annihilated by the operators
$P^*_i(v)$ and $P^\prime_{i\u}(v)$.  In the second case, the
$J^\P=1^+$ and $J^\P=2^+$ mesons are associated with $P_{i\u}(v)$ and
$P^*_{i\u\v}(v)$.  When the heavy quark is charm, we identify the individual
$SU(3)$ components of all these operators with the excited meson states as
follows:
\foot{In the absence of a universally accepted nomenclature for P-wave mesons,
we adopt the convention of labeling states with total angular momentum
subscripts and electric charge superscripts.  In the course of the text, we
also frequently follow the common albeit informal practice of denoting P-wave
mesons with double asterisk superscripts.}
\eqn\mesonstates{\eqalign{
(P_1^*, P_2^* P_3^*) &= (D_0^0, D_0^+, D_{s0}) \cr
(P'_{1\u},P'_{2\u},P'_{3\u}) &= (D_1^{0'},D_1^{+'},D'_{s1}) \cr
(P_{1\u},P_{2\u},P_{3\u}) &= (D_1^0, D_1^+, D_{s1}) \cr
(P_{1 \u\v}^*, P_{2\u\v}^*, P_{3\u\v}^*) &= (D_2^0, D_2^+, D_{s2}). \cr}}
The scalar $P_i^*$ and axialvector $P'_{i\u}$ operators may be assembled into
the superfield
\eqn\Sfield{S_i(v) = \proj \bigl[ - P^*_i(v) + P^\prime_{i \u}(v) \gamma^\u
\gfive \bigr] }
which is just the parity reversed analog of $H$.  The axialvector $P_{i\u}$
and traceless, symmetric tensor $P_{i\u\v}^*$ operators may be combined into
a third superfield \FalkLuke
\eqn\Tfield{T_i^\u = \proj \bigl[ P^{*\u\v}_i(v) \gamma_\v - \sqrt{3 \over 2}
P_{i\v}(v) \gfive \bigl( g^{\u\v} - \third \g^\v [ \g^\u-v^\u] \bigr) \bigr].}
$S$ and $T^\u$ transform exactly like $H$ under $SU(3)_{\L+\R}$ and $SU(2)_v$.

	It is important to note that $H$, $S$ and $T^\u$ obey the following
constraints which restrict the form of interactions that one can write down
for these fields:
\eqna\superfieldconstraints
$${\qquad\qquad\qquad
\eqalign{\proj H(v) &= H(v) \cr
	    \proj S(v) &= S(v) \cr
	    \proj T^\u(v) &= T^\u(v) \cr} \qquad\qquad
\eqalign{H(v) \projminus &= H(v) \cr
	 S(v) \proj &= S(v) \cr
	 T^\u(v) \projminus &= T^\u(v). \cr}
\qquad\qquad\qquad\qquad
\eqalign{& \superfieldconstraints a \cr
	 & \superfieldconstraints b \cr
	 & \superfieldconstraints c \cr}} $$
Multiplication on the left by the projection operator $P_+=(1+\slash{v})/2$
simply picks out the two heavy quark degrees of freedom in all the
meson superfields.  Multiplication on the right by $P_\pm$ effectively
projects out two light degrees of freedom.  These two conditions account for
a total of four degrees of freedom within $H$ and $S$ corresponding to one
$J=0$ and three $J=1$ meson states.  The $T^\u$ superfield obeys two
additional auxiliary constraints \Falk
\eqn\Tconstraints{v_\u T^\u(v) = T^\u(v) \g_\u = 0}
that reduce its degrees of freedom to eight.  $T^\u$ thus precisely
accommodates three $J=1$ and five $J=2$ states.

	Interactions involving the meson superfields are further constrained by
reparameterization invariance \refs{\WiseII,\LukeManohar}.  Recall that
the decomposition $p=Mv+k$ of a heavy particle's four-momentum in terms of
its four-velocity $v$ and residual momentum $k$ is somewhat arbitrary.  To
$O(1/M)$, no physical result should be altered if these parameters are
redefined as
\eqn\paramshifts{\eqalign{v &\to v+\e/M \cr
			  k &\to k-\e \cr}}
where $v \cdot \e = 0$.  This change of variables leaves the total
four-momentum $p$ invariant and induces only an $O(1/M^2)$ correction to
$v^2=1$.  In Heavy Hadron Chiral Perturbation Theory, the parameter
redefinitions generate shifts in the meson superfields
\eqn\supershifts{\eqalign{H &\to H+\d H \cr
			  S &\to S+\d S \cr
			  T^\u &\to T^\u + \d T^\u \cr}}
which are fixed by the constraints in \superfieldconstraints{}\ and
\Tconstraints{}\ and by a superfield normalization condition.
For instance, varying the relations $\slash{v} H = H$ and $H \slash{v} = -H$
which follow from \superfieldconstraints{a}\ yields
\eqn\transformedconstraints{\eqalign{
\bigl(\slash{v}+{\slash{\e}\over M}\bigr)\bigl(H+\d H \bigr) &=
  H+\d H \cr
\bigl(H+\d H \bigr) \bigl(\slash{v}+{\slash{\e}\over M}\bigr)&=
 -\bigl(H+\d H\bigr). \cr}}
Solving these equations along with the normalization constraint
$\Tr(\bar{H} H) = \Tr\bigl[ (\bar{H}+\d\bar{H})(H+\d H) \bigr]$
for $\d H$, one readily deduces that it is proportional to a commutator:
\eqn\Hshift{\d H = \Bigl[ {\slash{\e}\over 2M}, H \Bigr].}
In a similar manner, we find the variations in the $S$ and $T^\u$
superfields
\eqn\STshifts{\d S = \Bigl\{ {\slash{\e}\over 2M}, S \Bigr\}
\quad{\rm and} \quad
\d T^\u = \Bigl[ {\slash{\e}\over 2M}, T^\u \Bigr]
-{\e_\v T^\v \over M} v^\u.}
The requirement that the effective theory remain invariant under the
transformations in \paramshifts\ and \supershifts\ then forbids certain terms
such as $\Tr(\bar{H} i D_\u S \g^\u)$ and $\Tr(\bar{H} iD_\u T^\u)$ from
appearing in the chiral Lagrangian \FalkLuke.

	With the meson superfields in hand, one can readily write down
the leading order effective chiral Lagrangian that describes the low energy
interactions between Goldstone bosons and mesons in the infinite heavy
quark mass limit.  The lowest order terms must be hermitian, Lorentz
invariant and parity even.   They must also respect the light chiral and
heavy quark spin symmetries and be consistent with reparameterization
invariance:
\eqn\Lzero{\eqalign{\CL^{(0)}_v &= \sum_{Q=c,b} \Bigl\{
  -\Tr \bigl[\, \bar{H} i v \cdot D H \,\bigr]
  +\Tr \bigl[\, \bar{S} (i v \cdot D - \DM_\S) S \,\bigr]
  +\Tr \bigl[\, \bar{T}_\u (i v \cdot D - \DM_\T) T^\u \,\bigr] \cr
  & \qquad + g_1 \Tr \bigl[\, \bar{H} H \slash{{\bf A}} \gfive \,\bigr]
	  + g_2 \Tr \bigl[\, \bar{S} S \slash{{\bf A}} \gfive \,\bigr]
	 + g_3 \Tr \bigl[\, \bar{T}_\u T^\u \slash{{\bf A}} \gfive \,\bigr] \cr
  & \qquad + f_1 \Tr \bigl[\, \bigl(\bar{H} S + \bar{S} H \bigr)
     v \cdot {{\bf A}} \gfive \,\bigr]
	   + f_2 \Tr \bigl[\, \bigl( \bar{S} T^\u + \bar{T}^\u S \bigr)
     {{\bf A}}_\u  \gfive \,\bigr]
  \Bigr\}.  \cr}}
The splittings $\DM_\S = M_\S - M_\H$ and $\DM_\T = M_\T - M_\H$ between the
excited and ground state multiplets are independent of heavy quark flavor and
do not vanish in the infinite quark mass limit.  They are consequently
included into the kinetic part of the zeroth order Lagrangian.  In the
interaction terms, the Goldstone fields couple to the mesons through
the axial vector combination ${\bf A}^\u = i(\xi^\dagger \partial^\u \xi-
\xi \partial^\u \xi^\dagger)/2$.  They also communicate via the vector field
${\bf V}^\u = (\xi^\dagger \partial^\u \xi + \xi \partial^\u \xi^\dagger)/2$
which resides within the covariant derivatives
\eqn\heavyderivs{\eqalign{
D^\u H &= \partial^\u H - H ({\bf V}^\u) \cr
D^\u S &= \partial^\u S - S ({\bf V}^\u) \cr
D^\u T^\v &= \partial^\u T^\v - T^\v ({\bf V}^\u). \cr}}

	The heavy meson chiral Lagrangian may be used to study strong
interaction transitions among states within the $H$, $S$ and $T^\u$
superfields.  We will focus upon the decays of P-wave $D_s^{**}$ mesons in
the following sections.

% ----------------------------------------------------------------------
\newsec{Kaon Decays of $D_s^{**}$ Mesons}
% ----------------------------------------------------------------------

	The simplest $D_s^{**}$ decay processes involve emission of a single
Goldstone boson which must emerge in an even partial wave to conserve parity.
Single pion decay of these $I=0$ states violates isospin, and
eta decay is either severely phase space suppressed or kinematically
forbidden.  The single Goldstone boson which these strange charmed mesons
therefore mainly eject is a kaon whose mass is comparable to the splittings
between the excited and ground state multiplets.

	If kinematically allowed, the strange members of the $S$ multiplet
predominantly decay through an $\ell=0$ partial wave down to states in $H$ via
the term proportional to $f_1$ in \Lzero.  As guaranteed by heavy quark spin
symmetry \IsgurWise, the rates for the two charged kaon modes
\eqna\Skaonrates
$$ \eqalignno{
\Gamma\bigl(D_{s0} \to D^0 K^+ \bigr) &= {f_1^2 \over 4 \pi}
\Bigl( {M_{D^0} \over M_{D_{s0}}} \Bigr) {E_K^2 | {\vec p}_K | \over f_K^2}
& \Skaonrates a \cr
\Gamma\bigl(D'_{s1} \to D^{*0} K^+ \bigr) &= {f_1^2 \over 4 \pi}
\Bigl( {M_{D^{*0}} \over M_{D'_{s1}}} \Bigr) {E_K^2 | {\vec p}_K | \over f_K^2}
& \Skaonrates b \cr}$$
are equal to lowest order in the $1/\mc$ expansion.  The same is true for
the neutral kaon decays $D_{s0} \to D^+ K^0$ and $D'_{s1} \to D^{*+}
K^0$.  The equality among rates is broken however by formally subleading but
phenomenologically important spin and flavor symmetry violating effects.  We
therefore choose to input actual meson masses into the kaon energies
$E_K = M_{D^{**}_s} - M_{D^{(*)}}$ and three-momenta $ |{\vec p}_K| =
\sqrt{E_K^2-M_K^2}$ in \Skaonrates{}.  We similarly set the Goldstone boson
decay parameter $f$ equal to $f_\K=113 \MeV$ rather than
$f_\pi=93 \MeV$ in the kaon decay rates.

	Single Goldstone boson transitions between the $T^\u$ and $H$
multiplets must proceed through an $\ell=2$ partial wave to
conserve angular momentum.  None of the dimension-4 terms in the leading order
chiral Lagrangian can contribute to such processes. However at next-to-leading
order, there exists a unique dimension-5 operator which does mediate D-wave
decays:
\foot{There is no operator analogous to the one in (3.2) with the symmetric
Goldstone expression $D_\u {\bf A}_\v + D_\v {\bf A}_\u$ replaced by
$D_\u {\bf A}_\v - D_\v {\bf A}_\u$ since the antisymmetric combination
identically vanishes.  Consequently, there exists only one independent
dimension-5 operator which mediates $T^\u \to H \pib$ and not two as
claimed in ref.~\FalkLuke.}
\eqn\Lone{\CL^{(1)}_v = \sum_{Q=c,b} \Bigl\{ {ih \over \LX}
  \Tr \bigl[\, \bigl( \bar{H} T^\u + \bar{T}^\u H \bigr) \g^\v \gfive \,\bigr]
  \bigl( D_\u {\bf A}_\v + D_\v {\bf A}_\u \bigr) + \cdots \Bigr\}.}
Using the spin sums
\eqn\spinsums{\eqalign{
\sum_{\I=1}^3 {\e^{(\I)}_{\u}}(v)^* \e^{(\I)}_{\v}(v) = &- g_{\u\v}+v_\u
v_\v\cr
\sum_{\I=1}^5 {\e^{(\I)}_{\u\v}}(v)^* \e^{(\I)}_{\a\b}(v) =
&- \third (g_{\u\v}-v_\u v_\v)(g_{\a\b}-v_\a v_\b) \cr
&+ \half (g_{\u\a}-v_\u v_\a)(g_{\v\b}-v_\v v_\b)
 +\half (g_{\u\b}-v_\u v_\b) (g_{\v\a}-v_\v v_\a)}}
to average and sum over initial and final state polarizations, one finds
the following rates for the allowed $T^\u \to H K$ transitions \FalkLuke:
\eqna\Tkaonrates
$$ \eqalignno{
\Gamma\bigl(D_{s1} \to D^* K \bigr) &= {5 \over 15 \pi}
 \Bigl({h \over f_K \LX}\Bigr)^2 \Bigl( {M_{D^*} \over M_{D_{s1}}} \Bigr)
 |{\vec p}_K|^5 & \Tkaonrates a \cr
\Gamma\bigl(D_{s2} \to D K \bigr) &= {2 \over 15 \pi}
 \Bigl({h \over f_K \LX}\Bigr)^2 \Bigl( {M_D \over M_{D_{s2}}} \Bigr)
 |{\vec p}_K|^5 & \Tkaonrates b \cr
\Gamma\bigl(D_{s2} \to D^* K \bigr) &= {3 \over 15 \pi}
 \Bigl({h \over f_K \LX}\Bigr)^2 \Bigl( {M_{D^*} \over M_{D_{s2}}} \Bigr)
 |{\vec p}_K|^5. & \Tkaonrates c \cr}$$
As predicted by general spin symmetry arguments. these results occur in
the ratio 5:2:3 in the infinite charm mass limit.

	The coupling constant $h$ multiplying the dimension-5 operator
in \Lone\ can be fixed from the decay rates of the $J^\P=2^+$ flavor partners
of $D_{s2}$.  Their single pion and eta widths are simply related by $SU(3)$
to the kaon expressions in \Tkaonrates{b,c}.  We set the sum of the rates
for $D_2^0$ $\to$ $D^+ \pi^-$, $D^0 \pi^0$, $D^0 \eta$, $D^{*+} \pi^-$ and
$D^{*0} \pi^0$ equal to the total width $\Gamma(D_2^{0}) = 28^{+8+6}_{-7-6}
\MeV$ recently reported by CLEO \CLEOmesonref.  Solving for $h$ then
yields the reasonable coupling constant value
$h/\LX = (0.23 \pm 0.04) /1000 \MeV.$  Once $h$ is fixed, we can predict the
width of $D_{s2}$ as a function of its mass by summing the partial widths for
$D_{s2}$ $\to $ $D^0 K^+$, $D^+ K^0$, $D_s \eta$, $D^{*0} K^+$ and $D^{*+}
K^0$.  The results for the central value and one-standard deviation of
$\Gamma(D_{s2})$ are plotted in \Dstwowidth.  As can be seen in the figure,
the $D_{s2}$ width lies in the 5-15 MeV range.

	 We should comment upon the uncertainties associated with our width
prediction. The two basic ingredients that have gone into this result are
$SU(3)$ and heavy quark spin symmetry.  $SU(3)$ relates
$\Gamma(D^0_2 \to D \pi) $ to $\Gamma(D_{s2} \to D K)$ and
$\Gamma(D^0_2 \to D^* \pi)$ to $\Gamma(D_{s2} \to D^* K)$.   Spin symmetry on
the other hand relates the partial widths $\Gamma(D^0_2 \to D K)$ and
$\Gamma(D^0_2 \to D^* K)$.   The CLEO collaboration has recently reported an
updated measurement for the ratio of the $SU(3)$ analogues of these last two
decay rates \CLEOmesonref:
\eqn\ratio{R={\Br (D_2^0 \to D^+ \pi^-) \over \Br (D_2^0 \to D^{+*}
\pi^-)} = 2.1^{+0.6+0.6}_{-0.6-0.6}.}
The HHCPT value $R=2.2$ for this ratio agrees remarkably well with the CLEO
measurement and bolsters one confidence in the effective theory.  Moreover
if we simply adopt the experimental number in \ratio, then we do not in fact
need to invoke spin symmetry to predict $\Gamma(D_{s2})$.  Our width
result consequently only depends upon $SU(3)$.  Since this flavor symmetry
is generally violated at the $30 \%$ level, we expect corrections of this
order to $\Gamma(D_{s2})$ beyond the one-standard deviation shown in
\Dstwowidth\ which originates from the experimental uncertainty in
$\Gamma(D_2^0)$.  A more precise estimate for the magnitude of $SU(3)$
corrections could in principle be determined by working at next-to-leading
order in the chiral expansion.  But unfortunately,  a full subleading order
analysis would introduce so many new operators with unknown coefficients that
all predictive power would be lost.

	We next turn to consider single Goldstone boson decays of axialvector
P-wave mesons which are complicated by mixing among the $J^\P=1^+$ states in
the $S$ and $T^\u$ multiplets.  In the infinite heavy quark mass limit, the
superfields' light
angular momenta quantum numbers $j_\ell = 1/2$ and $j_\ell = 3/2$ are exact
and preclude intermultiplet mixing.  However for $\mQ \ne \infty$, the
physical mass eigenstates are annihilated by linear combinations
\eqn\masseigenstates{\eqalign{
Q^\L_{i\u} &= \cos\th P'_{i\u} + \sin\th P_{i\u} \cr
Q^\H_{i\u} &= -\sin\th P'_{i\u} + \cos\th P_{i\u} \cr}}
of the axialvector operators inside $S$ and $T^\u$.  We identify the observed
$D_1^0$, $D_1^+$ and $D_{s1}$ mesons with $Q_1^\H$, $Q_2^\H$ and
$Q_3^\H$ respectively.

	$D_{s1}$ is the only excited strange charmed meson which has been
seen so far.  Its width however has not yet been experimentally resolved,
and only a $90\%$ CL upper bound $\Gamma~\le~2.3~\MeV$ on its total decay
rate has been set \CLEOmesonref.  We can use this limit to restrict the
magnitude of the S-wave component to the meson's width.  The constraint
translates into an upper bound on the coefficient $f_1$.  The results are
listed in Table~I for three different values of the mixing angle $\th$:
%
%\vfill\eject
%
$$ \vbox{\offinterlineskip
\def\tablerule{\noalign{\hrule}}
\hrule
\halign {\vrule#& \strut#&
\ \hfil#\hfil& \vrule#&
\ \hfil#\hfil& \vrule#&
\ \hfil#\hfil& \vrule#&
\ \hfil#\hfil\ & \vrule# \cr
\tablerule
height10pt && \omit && \omit && \omit && \omit &\cr
&& \quad $\theta$ \quad && \quad $\Gamma_\S(Q_3^\H)_{\rm max}\, /
 \,{\rm MeV}$ \quad && \quad $\Gamma_\D(Q_3^\H)\, / \,{\rm MeV}$ \quad &&
 \quad $(f_1)_{\rm max}$ \quad & \cr
height10pt && \omit && \omit && \omit && \omit &\cr
\tablerule
height10pt && \omit && \omit && \omit && \omit &\cr
&& $1^\circ$ && $2.166$ && $0.134$ && $3.82$ &\cr
height10pt && \omit && \omit && \omit && \omit &\cr
&& $5^\circ$ && $2.167$ && $0.133$ && $0.77$ &\cr
height10pt && \omit && \omit && \omit && \omit &\cr
&& $10^\circ$ && $2.170$ && $0.130$ && $0.38$ &\cr
height10pt && \omit && \omit && \omit && \omit &\cr
\tablerule}} $$
\centerline{Table I}
\medskip\noindent
As can be seen in the last column, the maximum limits on $f_1$ are all
of order unity.  They are thus consistent with one's general expectations
for a dimensionless coupling appearing in the leading order chiral Lagrangian.

	With $f_1$ constrained, we can use $SU(3)$ symmetry to bound the ratio
of S and D partial widths for the nonstrange $D_1^0$ meson.  In the past,
mixing between the $J^\P=1^+$ states in the $S$ and $T^\u$ multiplets
has been thought to induce a large S-wave component to the total width of the
physical $D_1^0$ mass eigenstate.  Since the S-wave decay rate is
significantly greater than the D-wave's, even a small $\ell=0$ admixture was
believed to generate a sizable width enhancement and to account for most of the
measured total width $\Gamma(D_1^0)=20^{+6+3}_{-5-3} \MeV$. (``A small
grapefruit can be larger than a typical apple'' \Lu.) Our bounds however for
the S/D ratio lead to the opposite conclusion. (``A tiny grapefruit is smaller
than a typical apple.'')  As can be seen from the entries in Table~II, the
$\ell=2$ component of the total width actually dominates over the $\ell=0$
contribution for reasonable values of $\th$:
\foot{The CLEO and E687 collaborations have reported that they see no
evidence for any S-wave contribution to $D_1^0$ decay
\refs{\CLEOmesonref,\Fermigrp}.  Their conclusion relies however upon the
questionable assumption of no final state interactions.}
$$ \vbox{\offinterlineskip
\def\tablerule{\noalign{\hrule}}
\hrule
\halign {\vrule#& \strut#&
\ \hfil#\hfil& \vrule#&
\ \hfil#\hfil& \vrule#&
\ \hfil#\hfil& \vrule#&
\ \hfil#\hfil\ & \vrule# \cr
\tablerule
height10pt && \omit && \omit && \omit && \omit &\cr
&& \quad $\theta$ \quad && \quad $\Gamma_\S(Q_1^\H)_{\rm max}\, /
 \,{\rm MeV}$ \quad && \quad $\Gamma_\D(Q_1^\H)\, / \,{\rm MeV}$ \quad
 && \quad $\Gamma_{\rm tot}(Q_1^\H)_{\rm max}\, / \,{\rm MeV}$ \quad & \cr
height10pt && \omit && \omit && \omit && \omit &\cr
\tablerule
height10pt && \omit && \omit && \omit && \omit &\cr
&& $1^\circ$ && $3.35$ && $7.41$ && $10.76$ &\cr
height10pt && \omit && \omit && \omit && \omit &\cr
&& $5^\circ$ && $3.36$ && $7.35$ && $10.71$ &\cr
height10pt && \omit && \omit && \omit && \omit &\cr
&& $10^\circ$ && $3.36$ && $7.19$ && $10.54$ &\cr
height10pt && \omit && \omit && \omit && \omit &\cr
\tablerule}} $$
\centerline{Table II}
\bigskip\noindent
The S-wave component therefore does not explain the $2\sigma$ discrepancy
between theory and experiment for the total $D_1^0$ decay rate.  The source
of this disagreement remains poorly understood.

% ----------------------------------------------------------------------
\newsec{Pion Decays of $D_s^{**}$ Mesons}
% ----------------------------------------------------------------------

	The rates for single kaon decay of $D_s^{**}$ mesons depend critically
upon the precise splittings among all the strange states in $H$, $S$ and
$T^\u$.  At this time, only the mass $M=2535.1 \pm 0.6 \MeV$ of $D_{s1}$ has
been measured \CLEOmesonref.  Assuming that its $J^\P=2^+$ partner has an equal
or greater mass, both of the states associated with $T^\u$ predominantly
decay via kaon transitions.  The situation for the strange members of the $S$
multiplet however is less clear and more interesting since their masses are
unknown.  Several attempts have been made to estimate the energy levels of
$D_{s0}$ and $D'_{s1}$ using the quark model.  $D_{s0}$ is generally
predicted to be heavy enough so that kaon decay is kinematically possible.
On the other hand, the variation in results from different quark model
calculations for the $D'_{s1}$ mass is sufficiently great that one cannot
conclusively determine whether single kaon decay is allowed \QMrefs.  It is
consequently possible that this strange meson must decay via other channels.

	The first alternative to consider is the isospin violating
transition $D'_{s1} \to D_s^* \pi^0$.  The basic characteristics of this
mode are very similar to those for $D_s^* \to D_s \pi^0$ which has recently
been studied in ref.~\ChoWise.  It proceeds at tree level through emission of
a virtual eta that subsequently mixes into a neutral pion.  The intermediate
$\eta$ propagator effectively renders the amplitude inversely proportional to
the strange quark mass which is smaller than a typical hadronic scale.
The isospin violation factor associated with $D'_{s1} \to D_s^* \pi^0$
\GasserLeutwyler
\eqn\quarkratio{(m_d - m_u) / \bigl(m_s - (m_u+m_d)/2 \bigr) \simeq 1/43.7}
is therefore not so suppressed as one might have thought.  Setting
the Goldstone boson decay parameter $f$ equal to $f_\eta = 121 \MeV$, we find
the rate
\eqn\isodecay{\Gamma(D'_{s1} \to D_s^* \pi^0) = {f_1^2 \over 32 \pi}
\Bigl( {M_{D_s^*} \over M_{D'_{s1}}} \Bigr)
\Bigl[ {m_d - m_u \over m_s - (m_u+m_d)/2} \Bigr]^2
{E_\pi^2 |{\vec p}_\pi| \over f_\eta^2}}
for the single pion process.

	Isospin conserving double pion emission represents the next most
important decay mode for strange charmed P-wave mesons.  Unlike their
nonstrange counterparts \FalkLuke, $I=0$ $D_s^{**}$ mesons cannot
undergo double pion decay via pole graphs in which two $I=1$ pions are
sequentially emitted.  The two pions must instead emerge in an isospin zero
combination from higher order terms in the chiral Lagrangian.  At
dimension-5, there exist just two such operators which mediate the superfield
transitions $S \to H \pi \pi$ and $T^\mu \to H \pi \pi$ and preserve heavy
quark spin symmetry:
\eqn\moreLone{\CL^{(1)}_v = \sum_{Q=c,b} \Bigl\{
 {h_1\over\LX} \Tr\bigl[\,\bigl(\bar{H} S  + \bar{S} H \bigr) v^\u \g^\v
  \bigl] \Tr [{\bf A}_\u {\bf A}_\v ]
+ {h_2 \over \LX} \Tr \bigl[\, \bigl( \bar{H} T^\u + \bar{T}^\u H \bigr)
  v^\v \,\bigr] \Tr [{\bf A}_\u {\bf A}_\v] + \cdots \Bigr\}.}
At $O(1/\mc)$, additional spin symmetry violating operators can participate
as well.  We will focus however upon the leading terms in \moreLone.

	Working in the two pion center of mass frame, we can readily decompose
the $D_s^{**} \to D_s^{(*)} \pi \pi$ decay amplitudes into S-wave and
D-wave components.  After squaring the amplitudes, we find the
following $S \to H \pi \pi$ differential decay rates for these partial waves
and their interference term:
\eqna\Sdiffrates
$$ \eqalignno{
d\Gamma \bigl( D_{s0} \to D_s^* \sum_{i=1}^3 \pi^i \pi^i \bigr)_{\S,\D,\S\D}
  &= {3 \over 2} \Bigl( {h_1 \over f_\pi^2 \LX} \Bigr)^2
  (F_{\S,\D,\S\D}) d \Phi_3 & \Sdiffrates a \cr
d\Gamma \bigl( D'_{s1} \to D_s \sum_{i=1}^3 \pi^i \pi^i \bigr)_{\S,\D,\S\D} &=
  {3 \over 2} \Bigl( {h_1 \over f_\pi^2 \LX} \Bigr)^2
  \bigl({1 \over 3} F_{\S,\D,\S\D} \bigr) d \Phi_3& \Sdiffrates b \cr
d\Gamma \bigl( D'_{s1} \to D_s^* \sum_{i=1}^3 \pi^i \pi^i \bigr)_{\S,\D,\S\D}
  &= {3 \over 2} \Bigl( {h_1 \over f_\pi^2 \LX} \Bigr)^2
  \bigl({2 \over 3} F_{\S,\D,\S\D} \bigr) d \Phi_3.
  & \Sdiffrates c \cr} $$
For completeness, we also quote the corresponding $T^\mu \to H \pi \pi$
differential widths even though they are in reality very small compared to
the $T^\u \to H K$ rates in \Tkaonrates{}:
\eqna\Tdiffrates
$$ \eqalignno{
d\Gamma \bigl( D_{s1}  \to D_s \sum_{i=1}^3 \pi^i \pi^i \bigr)_{\S,\D,\S\D} &=
  {3 \over 2} \Bigl( {h_2 \over f_\pi^2 \LX} \Bigr)^2
  \bigl({2 \over 9}  F_{\S,\D,\S\D} \bigr) d \Phi_3 & \Tdiffrates a \cr
d\Gamma \bigl( D_{s1} \to D^*_s \sum_{i=1}^3 \pi^i \pi^i \bigr)_{\S,\D,\S\D} &=
  {3 \over 2} \Bigl( {h_2 \over f_\pi^2 \LX} \Bigr)^2
  \bigl({1 \over 9} F_{\S,\D,\S\D} \bigr) d \Phi_3 & \Tdiffrates b \cr
d\Gamma \bigl( D_{s2} \to D_s^* \sum_{i=1}^3 \pi^i \pi^i \bigr)_{\S,\D,\S\D} &=
  {3 \over 2} \Bigl( {h_2 \over f_\pi^2 \LX} \Bigr)^2
  \bigl({3 \over 9} F_{\S,\D,\S\D} \bigr) d \Phi_3. & \Tdiffrates c \cr} $$
As required by heavy quark spin symmetry, the differential widths in
\Sdiffrates{a,b,c}\ and \Tdiffrates{a,b,c}\ occur in the ratios $3:1:2$ and
$2:1:3$ respectively in the infinite charm mass limit \IsgurWise.

	The functions $F_\S$, $F_\D$ and $F_{\S\D}$ entering into
\Sdiffrates{}\ and \Tdiffrates{}\ may be conveniently expressed in a
manifestly Bose symmetric form in terms of the energies $E_1$ and $E_2$ of
the two pions measured in the decaying $D_s^{**}$ rest frame, the
pion pair invariant mass $s$, and the mass splitting $\DM$ between the
initial and final heavy mesons:
\eqn\Ffunctions{\eqalign{
F_\S &= \ninth {\DM^2 \over s^2} (\DM^2-s) (s+2\mpi^2)^2 \cr
F_\D &= \quarter \Bigl[ (\DM^2-4E_1 E_2)^2 + \bigl(1-{2 \over 3}
{\DM^2 \over s} \bigr) \bigl(\DM^2-4 E_1 E_2 \bigr) (s - 4\mpi^2) \cr
&\qquad +\ninth {\DM^2 \over s^2} (\DM^2 - s) (s-4\mpi^2)^2 \Bigr] \cr
F_{\S\D} &= -\third {\DM^2 \over s} (s+2 \mpi^2) \Bigl[\DM^2-4E_1 E_2
-\third \bigl( {\DM^2 \over s} - 1 \bigr) (s-4\mpi^2) \Bigr]. \cr}}
The three body phase space factor can also be simply written in terms of these
variables:
\eqn\phasespace{d \Phi_3 = {1 \over 64 \pi^3} \d(\DM-E_1-E_2) \, d E_1 \,
d E_2 \, ds.}
Recall that in a charm or bottom hadron decay, the heavy body in the final
state must generally recoil in order to conserve momentum.  However, it
carries away no kinetic energy in the limit of its mass tending towards
infinity.  The energy conserving delta function in $d \Phi_3$
therefore constrains the two final state pions to take away all of the
kinetic energy released by the original $D_s^{**}$ meson.

	The double pion decay rates in \Sdiffrates{}\ and \Tdiffrates{}\ can
be used to test basic HQET ideas.  The extent to which these differential
widths will agree or disagree with future experimental measurements provides
some measure of the importance of $O(1/\mc)$ spin-flavor violating effects.
Since the presumption that the charm quark is truly heavy represents the
weakest point in most HQET applications, it is important to test this
hypothesis in as many different settings as possible.  Our P-wave meson decay
expressions provide such an opportunity.

	The differential two pion rates may be integrated to obtain the
corresponding total partial wave widths.  Integrating the functions in
\Ffunctions\ over $s$ between its upper and lower limits for fixed $E_1$ and
$E_2$
\eqn\slimits{s_\pm = 2 \bigl[ \mpi^2 + E_1 E_2 \pm \sqrroot \bigr],}
we find
\eqn\integFfunctions{\eqalign{
\int^{s_+}_{s_-} ds F_\S &= {4 \over 9} \DM^2 \Bigl\{ \bigl[ \DM^2 - 5 \mpi^2
 - 2 E_1 E_2 \bigr] \sqrroot \cr
 &\qquad\qquad + \mpi^2 (\DM^2-\mpi^2) \log{\mpi^2+E_1 E_2 +
 \sqrroot \over \mpi^2+E_1 E_2 - \sqrroot} \Bigr\} \cr
\int^{s_+}_{s_-} ds F_\D &= {4 \over 9} \DM^2 \Bigl\{ \bigl[ \DM^2 - 2 \mpi^2
 + \bigl( 18 {\mpi^2 \over \DM^2}-8 \bigr) E_1 E_2 + 18 {(E_1 E_2)^2 \over
 \DM^2} \bigr] \sqrroot \cr
 &\qquad\qquad + \mpi^2 \bigl( \DM^2-\mpi^2-6 E_1 E_2 \bigr)
 \log{\mpi^2+E_1 E_2 + \sqrroot \over \mpi^2+E_1 E_2 - \sqrroot} \Bigr\} \cr
\int^{s_+}_{s_-} ds F_{\S\D} &= -{8 \over 9} \DM^2 \Bigl\{ \bigl[ \DM^2 +
 \mpi^2 - 5 E_1 E_2 \bigr] \sqrroot \cr
 &\qquad\qquad + \mpi^2 (\DM^2-\mpi^2-3 E_1 E_2)
 \log{\mpi^2+E_1 E_2 +  \sqrroot \over \mpi^2+E_1 E_2 - \sqrroot} \Bigr\}.\cr}}
We perform the remaining integrations over pion energies numerically.

	We should note that final state interactions have been neglected
here.  The OZI rule leads one to expect that such interactions between the
strange heavy meson and the two pions are small.  Furthermore, experimentally
measured pion scattering phase shifts indicate that the interactions between
the pions themselves are also small in the D-wave.  However, S-wave pion final
state interactions are known to be important over much of the kinematic
regime for $D_s^{**} \to D_s^{(*)} \pi \pi$ transitions.
So although we expect our S-wave differential decay rates
to be accurate for invariant dipion masses near $s=4 \mpi^2$, its integrated
rate cannot really be trusted to provide much more than an order of magnitude
estimate.  Final state interactions could be incorporated into our results
by altering the S-wave amplitude calculated here to include the pion
scattering phase shift in a manner consistent with unitarity \Jackson.

	We plot the total integrated double pion width for $D'_{s1}$ as a
function of its mass in \Gammaplot.  The meson's single kaon and single pion
decay rates are also illustrated for comparison.  We have set the unknown
coupling constant $f_1$ which enters into $\Gamma(D'_{s1} \to D^* K)$ and
$\Gamma(D'_{s1} \to D_s^* \pi^0)$ equal to unity.  On the other hand, we have
set $h_1=0.1$ in $\Gamma(D'_{s1} \to D_s^{(*)} \sum_i \pi^i \pi^i)$
since it is OZI suppressed.  Given the substantial uncertainties
in these couplings, the curves in \Gammaplot\ provide only qualitative
information.  However, it is obvious from the figure that the kaon transition
completely dominates over the pion processes if it is kinematically allowed.
This is not surprising since the two body kaon decay is mediated by a
dimension-4 operator in the leading order chiral Lagrangian whereas the single
pion mode violates isospin while the double pion transition involves three
bodies in the final state and proceeds only at next-to-leading order.  It
is also likely that the $D'_{s1}$ state will never be observed if single
kaon emission is indeed kinematically possible since its width would be very
broad.  But if its mass turns out to be less than $2504 \MeV$, then the
$D'_{s1}$ width should be quite narrow.   In this situation, it will hopefully
be possible to observe this strange P-wave meson in the future through its
single or double pion channels.  Indeed, we believe this scenario represents
the best prospect for ever finding any of the $S$ multiplet mesons.

% ----------------------------------------------------------------------
\newsec{Conclusion}
% ----------------------------------------------------------------------

	Our study of Goldstone boson decays of strange charmed P-wave mesons
has yielded a number of results which can be experimentally tested.   Our
model independent prediction for the width of $D_{s2}$ lies in an
experimentally accessible range.  We are consequently optimistic that this
$J^\P=2^+$ state will soon be observed.   On the other hand, detecting
the $D'_{s1}$ partner of $D_{s1}$ is much more uncertain.  Single kaon
emission must be kinematically forbidden in order for this $D'_{s1}$ state
to ever be seen.   We believe however that it is worthwhile to search for this
resonance through its single and double two pion decay modes.

\bigskip
\centerline{\bf Acknowledgments}
\bigskip

      We thank John Bartelt, Jon Urheim and Mark Wise for several helpful
discussions.

\bigskip\bigskip\bigskip
\bigskip\bigskip

\noindent
{\it Note added:}  A few weeks after completion of this paper, the CLEO
collaboration reported the discovery of the $D_{s2}$.  Its measured mass and
width are $M_{D_{s2}} = 2573.2 \pm 1.9 \MeV$ and
$\Gamma_{D_{s2}} = 16^{+5+3}_{-4-3} \MeV$ \CLEODS.

\listrefs
\listfigs
\bye